\documentclass[12pt,preprint]{aastex}

\usepackage{psfig,natbib}   
\usepackage{emulateapj5}    

\newcommand{\mic}{\,$\mu$m }
\newcommand{\micpa}{\,$\mu$m}                                                   
\newcommand{\muJy}{\,$\mu$Jy }
                                                   
\bibpunct[]{(}{)}{;}{a}{}{,}

\shorttitle{Luminous Infrared Galaxies at $z \gtapp 1$}
\shortauthors{E.\,Le Floc'h et al.}

\accepted{2004 March 26}
\begin{document}
\def\gtapp
{\mathrel{\hbox{\raise0.3ex\hbox{$>$}\kern-0.8em\lower0.8ex\hbox{$\sim$}}}}
\def\ltapp
{\mathrel{\hbox{\raise0.3ex\hbox{$<$}\kern-0.75em\lower0.8ex\hbox{$\sim$}}}}
\def\ts{\thinspace}


\title{Identification of luminous infrared galaxies at 
$1 \ltapp$ z $\ltapp$ 2.5
$^{1,2,3,4,5,6}$}


\altaffiltext{1}{Based on observations made with {\it Spitzer}, 
 operated by the Jet Propulsion Laboratory under NASA
contract 1407.}  

\altaffiltext{2}{Based on observations collected at
the Subaru Telescope, which is operated by the National Astronomical
Observatory of Japan.} 

 \altaffiltext{3}{Based on observations
collected at the Canada-France-Hawaii Telescope, which is operated by
the National Research Council of Canada, the Centre National de la
Recherche Scientifique and the University of Hawaii.}

\altaffiltext{4}{Based on observations obtained at Kitt Peak National
Observatory, National Optical Astronomy Observatory, which is operated
by the Association of Universities for Research in Astronomy,
Inc. (AURA) under cooperative agreement with the National Science
Foundation.}

\altaffiltext{5}{Based on observations collected at the German-Spanish
Astronomical Center, Calar Alto, operated jointly by Max-Planck
Institut f\"{u}r Astronomie and Instituto de Astrofísica de Andalucia
(CSIC).}

\altaffiltext{6}{Based on observations made with the Isaac Newton 
Telescope, 
 operated on the island of La Palma by the Isaac Newton
Group in the Spanish Observatorio del Roque de los Muchachos of the
Instituto de Astrof\'{\i}sica de Canarias.}

\slugcomment{Accepted for publication in  The Astrophysical Journal Supplement,
May 6th, 2003}


\author{E.\,Le~Floc'h$^{\rm a}$, P.G.\,P\'erez-Gonz\'alez$^{\rm a}$,
G.H.\,Rieke$^{\rm a}$, C.\,Papovich$^{\rm a}$, J.-S.\,Huang$^{\rm b}$,
P.\,Barmby$^{\rm b}$, H.\,Dole$^{\rm a,c}$,
E.\,Egami$^{\rm a}$, A.\,Alonso-Herrero$^{\rm a}$, G.\,Wilson$^{\rm
d}$, S.\,Miyazaki$^{\rm e}$, J.R.\,Rigby$^{\rm a}$, L.\,Bei$^{\rm a}$,
M.\,Blaylock$^{\rm a}$, C.W.\,Engelbracht$^{\rm a}$, G.G.\,Fazio$^{\rm
b}$, D.T.\,Frayer$^{\rm d}$, K.D.\,Gordon$^{\rm a}$, D.C.\,Hines$^{\rm
a,f}$, K.A.\,Misselt$^{\rm a}$, J.E.\,Morrison$^{\rm a}$,
J.\,Muzerolle$^{\rm a}$, M.J.\,Rieke$^{\rm a}$, D.\,Rigopoulou$^{\rm
g}$, K.Y.L.\,Su$^{\rm a}$, S.P.\,Willner$^{\rm b}$ and
E.T.\,Young$^{\rm a}$} \affil{$^{\rm a}$Steward Observatory,
University of Arizona, Tucson, AZ 85721, USA \\ $^{\rm
b}$Harvard-Smithsonian Center for Astrophysics, 60 Garden Street,
Cambridge, MA 02138, USA \\ $^{\rm c}$Institut d'Astrophysique
Spatiale, Universit\'e Paris Sud, F-91405 Orsay Cedex, France \\
$^{\rm d}$Spitzer Science Center, Mail Code 220-6, 1200 East
California Boulevard, Pasadena, CA 91125, USA \\ $^{\rm e}$Subaru
Telescope, National Astronomical Observatory 650 N Aohoku Pl., Hilo
HI96720, USA \\ $^{\rm f}$Space Science Institute, 4750 Walnut Street,
Suite 205 Boulder, Colorado 80301, USA \\ $^{\rm g}$Astrophysics,
Denys Wilson Building, Keble Road, Oxford, OX1 3RH, UK}






\begin{abstract} 
We present preliminary results on 24\mic detections of luminous
infrared galaxies at $z \gtapp 1$ with the Multiband Imaging
Photometer for Spitzer (MIPS).
 Observations were performed in the
Lockman Hole and the Extended Groth Strip (EGS), and were supplemented
by data obtained with the Infrared Array Camera (IRAC) between 3
and 9\micpa. The positional accuracy of
$\ltapp$ 2\arcsec \, for most MIPS/IRAC detections provides
unambiguous identifications of their optical counterparts.
Using spectroscopic redshifts from the  Deep Extragalactic
Evolutionary Probe survey,
we identify
24\mic sources 
 at $z \gtapp 1$ in the EGS, while the combination of the MIPS/IRAC
observations with $BVRIJHK$ ancillary data in the Lockman Hole also
shows very clear cases of galaxies with photometric redshifts at $1
\ltapp z \ltapp 2.5$.

 The observed 24\mic fluxes 
indicate infrared luminosities greater than $10^{11}$\,L$_{\odot}$,
while the data at shorter wavelengths reveal rather red and probably
massive ($\mathcal{M}\gtapp\mathcal{M}^*$) galaxy counterparts.  It is
the first time that this population of luminous objects is detected up
to $z\sim2.5$
 in the infrared.
Our work demonstrates 
the ability of the MIPS instrument to probe
the dusty Universe at very high redshift, and illustrates how
the forthcoming {\it Spitzer\,} deep
surveys  will
offer a unique
opportunity to illuminate a dark side of cosmic history
not explored by previous infrared experiments.

\end{abstract}



\keywords{ galaxies: high-redshift ---  infrared: galaxies ---
 cosmology: observations}


\section{Introduction}


In the past few years, deep observations in the infrared and
submillimeter revealed a 
population of high redshift 
galaxies emitting the bulk of their luminosity
between 8 and 1000\micpa, whose cosmological significance
 had been
previously missed or severly underestimated by optical surveys
because of extinction effects due to dust. 
These sources,  likely the analogs of the local
Luminous and Ultra-Luminous InfraRed Galaxies (respectively
LIRGs: $10^{11}$\,L$_{\odot} \leq$ L$_{\rm IR} = $L$[8-1000\mu m] \leq 10^{12}$\,L$_{\odot}$,
and ULIRGs:  L$_{\rm IR} \geq 10^{12}$\,L$_{\odot}$)
played a crucial
role through the cosmic ages.
 At mid-infrared 
wavelengths, results from the {\it Infrared Space
Observatory\,} ($ISO$) mission show
that the evolution of these powerful objects noticeably contributed
to the global star formation history at $0 \ltapp z \ltapp 1$
 \citep{Elbaz99,Flores99}. 
Observations
 with SCUBA at 850\mic also led to the discovery
of extremely bright infrared sources mainly
located at $z \gtapp 2$
 and characterized by a
space density 1000 times larger than observed in the local
Universe 
\citep[e.g.,][]{Chapman03}.


From the purely observational point of view though,
 very little is currently known about the potential importance
of  dusty sources at
 $1 \ltapp z \ltapp 2$.
This is explained
on one hand by   sensitivity limitations of the 
infrared instruments previously used, such as ISOCAM and ISOPHOT on-board $ISO$, 
and on the other hand by the 
 SCUBA surveys  being mostly sensitive to
 L\,$\gtapp 10^{12}$L$_{\odot}$ galaxies 
which were much  more numerous
at  $z \sim$ 2-3.
Current knowledge indicates that 
a large fraction of today's stars were probably
born at $1 \ltapp z \ltapp 2$
 \citep[e.g.,][]{Dickinson03,Calura03}, before the global activity of
star formation began to  decrease rapidly down to $z \sim 0$
\citep{Lilly96}.


The recently commissioned Multiband Imaging Photometer for Spitzer
(hereafter MIPS) on-board {\it Spitzer\,} provides a unique
opportunity to address this issue. 
 For sources at $z \gtapp 1.5$, its  24\mic band
 is particularly suitable for detecting the possible redshifted
emission from the 8\mic broad-band feature commonly seen in infrared
galaxy spectra, while the main 15\mic filter on ISOCAM was rather more
adapted for detecting this feature at $z \sim 0.8$.
Here we 
assess the capabilities of the MIPS instrument to  
shed light on a yet unexplored side of galaxy evolution, and
present our
preliminary results on the detection of $1 \ltapp z \ltapp 2.5$
infrared galaxies at 24\micpa.
This work is one facet of a larger set of publications also addressing source number counts
at 24, 70 and 160\mic \citep{Papovich04,Dole04}, IR model interpretation
\citep{Lagache04}, as well as the relation between
 the MIPS-selected objects and other populations such as
 the SCUBA/MAMBO and VLA galaxies
\citep{Egami04,Ivison04,Serjeant04}
 and X-ray sources
 \citep{Alonso04,Rigby04}.
Throughout the paper, 
a $\Lambda$CDM cosmology with H$_0$\,=\,70~km~s$^{-1}$\,Mpc$^{-1}$,
$\Omega_m$\,=\,0.3 and $\Omega_{\lambda}\,=\,0.7$ is assumed.

\section{Observations and data analysis}

We performed 24\mic observations\footnote{{\it Spitzer\,} Prog.\,IDs 8
\& 1077.} with MIPS \citep{Rieke04} in a 5'$\times$5' area of the
Lockman Hole (LH, $\alpha=10^h51^m56^s$, $\delta=57^{\rm o}25'32''$,
J2000)
as well as in a 2$^{\rm o} \times$10' region of the
Extended Groth Strip (EGS, $\alpha \sim 14^h19^m$, $\delta \sim
52^{\rm o}48'$, J2000).
Data in the Lockman Hole were obtained
in the MIPS ``photometry mode'', with a
series of dithered exposures allowing a high frame redundancy and
efficient cosmic ray rejection \citep[see][ for more details]{Egami04}.
 Observations in the EGS were performed with the
MIPS  ``slow scan'' technique, an observing mode that
allows  coverage of large sky areas with high efficiency.
Data reduction was carried out using
the MIPS Data Analysis Tool 
\citep{Gordon04}.
Effective
 integration times of $\sim$\,250\,s and $\sim$\,450\,s per sky pixel
were reached respectively for the Lockman Hole and EGS data, leading
to a 24\mic 1$\sigma$ rms $\sim$\,35\muJy (LH) and $\sim$\,25\muJy (EGS).
 The astrometry of the Groth Strip final mosaic
was refined using bright
objects also listed in the 2MASS catalog \citep{Jarrett00}, 
while source positions in 
our smaller coverage of the Lockman Hole were matched
against an abundant set of ancillary data (see below). We
estimate the absolute
astrometry accuracy to be better than 2\arcsec.
Point source extraction and photometry were performed as
described by \citet{Papovich04}. 

Both fields were also
observed with the
{\it Spitzer InfraRed Array Camera} \citep[hereafter IRAC,][]{Fazio04}
at 3.6, 4.5, 5.8 and 8\micpa, down to 5$\sigma$~rms $\sim$1.2,
1.1, 6.5 and  8.0\muJy respectively. 
 A detailed description of the IRAC data 
analysis
is presented by \citet{Huang04}. In addition, we retrieved $HST$/WFPC2
images of the EGS from the public database of the Deep Extragalactic
Evolutionary Probe (DEEP) team\footnote{see \sf
http://deep.ucolick.org}, and a set of
ancillary observations of the Lockman Hole was collected in the
optical and the near-infrared (NIR) to supplement our {\it Spitzer\,}
program.  For the latter, we made use of archival $B$--band images from the
Wide Field Camera (WFC) of the Isaac Newton Telescope (INT), while
$VRIJHK_s$ data were obtained 
 with the UH8K camera on the
Canada--France--Hawaii Telescope ($VI$-bands), the
Suprime-CAM instrument on the Subaru Telescope ($R$-band), TIFKAM on
the 2.1m Telescope of Kitt Peak ($H$-band) and Omega-Prime on the 3.5m
Telescope of Calar Alto ($JK_s$-band).
Limiting Vega magnitudes
(3$\sigma$ level) of 26.0, 24.0, 27.0, 24.0, 22.0, 20.5,
20.0~mag\,arcsec$^{-2}$ were respectively derived in the $BVRIJHK_s$
filters.  Source detection and photometry were carried out with
Sextractor \citep{Bertin96}.

For the Lockman Hole data, the full-width at half maximum of the
PSF  goes from $\sim$1{\arcsec} in the optical/NIR bands to 2{\arcsec}
and 6$\arcsec$ in the IRAC and MIPS images, which
gave us  confidence in  identifying the respective
 counterparts at each wavelength (see an illustration in Fig.\,1).
Merged catalogues among the different filters
were then built,
 starting with matching
the coordinates of the 24\mic sources in the deep $R$-band image. 
Following \citet{Kron80}, we estimated the {\it total\,} flux
for each object detected in the $R$\,~image using an elliptical
aperture derived from an isophotal fit around the given source, 
and translated those apertures to
the other optical and NIR data to obtain integrated fluxes for the
corresponding filter (in all cases the apertures were large enough to
enclose the PSF profile).  IRAC and MIPS fluxes were derived
respectively with circular aperture photometry and PSF fitting
\citep[see][]{Huang04,Papovich04}.









\section{Results: Infrared galaxies at $z \gtapp 1$}

\subsection{Spectroscopic redshifts}

Our source position catalog in the EGS region was compared to the
DEEP1 database of galaxies with spectroscopic redshifts
(Weiner et al., in preparation).
Though most of objects from this survey
clearly lie at $0 \ltapp z \ltapp 1$, we identified from our 24\mic observations
nine
galaxies
at $1 \ltapp z \ltapp 1.3$, 
and two other sources at $z$=1.43 and
$z$=$1.74$ whose redshifts were respectively
derived from  H$\alpha$ \citep{Hopkins00} and
 Fe\,II absorption lines
 (B.\,Weiner, priv. communication).
From their observed 24\mic fluxes and
 using the luminosity-dependent spectral energy distribution (SED)
templates from the model of \citet{Lagache04}, we then estimated their
 infrared
luminosities between 8 and 1000\micpa. 
Even though the exact relation between the mid- and total IR galaxy
emission is not yet well understood, there are
some indications that the mid- to far-infrared correlation
still holds in the distant Universe 
\citep{Appleton04}, and the L$_{24\mu m}$/L$_{\rm IR}$
ratio anyhow varies by less than a factor of $\sim$2--3 for most model SEDs
at high reshifts \citep[e.g.,][]{Papovich02}. With this caution in
mind, we found 4~sources with 4$\times$10$^{11}$L$_{\odot}<$~L$_{\rm
IR} < 10^{12}$\,L$_{\odot}$, 6~galaxies with
10$^{12}$\,L$_{\odot}<$~L$_{\rm IR} <$
3$\times$10$^{12}$\,L$_{\odot}$, and one hyper luminous system
characterized by L$_{\rm IR}$=2.5$\times$10$^{13}$\,L$_{\odot}$.
Given that the objects from the DEEP1 database are optically bright,
we note that these luminous sources likely have higher optical/IR flux
ratios than those of more obscured infrared galaxies. This underlies the
importance of obtaining near-IR and mid-IR (e.g., IRS/Spitzer)
spectroscopy as well as  accurate photometric redshifts
to study all types of IR--luminous objects.

\subsection{Photometric redshifts}
\label{sec:temp}

 For distant galaxies, our analysis of SEDs for EGS sources with
spectroscopic redshifts indicates that the combination of the 4 IRAC
bands with optical and NIR data allows locating with good accuracy
the spectral feature produced both by the
stellar H$^-$ opacity minimum at
{\it rest-frame\,} 1.6\mic and the global shape of the 
 underlying stellar continuum.
This characteristic of galaxy SEDs provides a key indication for
photometric redshift determinations \citep[Huang et al. in
preparation, see also][]{Simpson99,Sawicki02}. 
Figure\,2a illustrates this result in the case of
two representative sources of our sample, whose 
broad-band SEDs 
lead to  photometric
redshifts in good agreement with those determined
from optical spectroscopy.


Relying on this new approach, we used our optical/NIR and IRAC data 
to locate 24\mic selected sources with photometric
redshifts $z \gtapp 1$.  
The analysis of the complete MIPS data set in the EGS is still
in progress. We present
in this preliminary study the results we obtained
for the Lockman Hole.
Over the 5'$\times$5' area covered by MIPS,
we identify {\it at least\,} 25 objects
characterized by 
a rather prominent stellar
bump revealing an unambiguous redshift larger than 1.
Figure\,2b shows the $BVRIJHK$/IRAC/MIPS broad-band
SEDs of several of these sources spanning the $1 \ltapp z \ltapp 2.5$
redshift range, with the best galaxy template fitted to our data.
These templates were built by combining the optical/NIR spectral
range emission
modeled by \citet{Devriendt99}  with the infrared SEDs from the model 
of \citet{Lagache04}. For several objects
showing a rather high MIPS to IRAC flux ratio, 
an additional contribution from a hot dust continuum 
(i.e., Very Small Grains)
was arbitrarily added \citep[see][ for more details on the
method]{Gallais04}.
In this first attempt to locate
$z \gtapp 1$ galaxies using the 1.6\mic stellar bump, we relied on 
a simple ``by-eye'' fit which clearly gives a sufficient constraint
(i.e., $\Delta$z$\sim$0.2) for the cases shown in Fig.\,2.
We emphasize that the few SEDs shown in Fig.\,2 were not chosen to be
globally representative of our galaxy sample at similar redshifts, but
correspond to 24\mic sources detected at least in the $R$/IRAC bands
and for which
a photometric
redshift could be reliably obtained from the spectral feature produced
at rest-frame 1.6\micpa.
This naturally biases the selection
against sources with a rather shallow stellar bump like
the active nuclei \citep[see][]{Alonso04}. This issue will obviously 
have to be
taken into account in future studies of more complete samples.

Table\,1 summarizes  general properties of our sample in the Lockman Hole.
As in the EGS, the estimated redshifts and  observed 24\mic fluxes
indicate
infrared luminosities 
characterized by 
L$_{\rm IR} > 10^{11}$\,L$_{\odot}$.
The rest-frame optical and NIR  properties 
 reveal rather red colors (i.e., observed $R-K>4$)
originating either from an evolved stellar population or
a young reddened starburst. They also 
indicate high luminosities
 in the
rest-frame near-infrared, which suggests the presence of
rather massive objects ($\mathcal{M}\gtapp\mathcal{M}^*$).

Finally, we  note that only one of these sources 
 is detected in the deep X-ray observations of the Lockman
Hole by XMM \citep{Hasinger01}. 
In conjunction with the galaxy-like
templates fitting their optical SEDs, the X-ray non-detections
 suggest that these galaxies
are mostly dominated by star-forming activity, 
even if the presence of an active nucleus
 can not be fully excluded by the data
\citep[see][]{Alonso04}.



\section{Discussion and Conclusion}

We stress that our reported number of 24\micpa-selected infrared
galaxies at $z \gtapp 1$ is obviously a lower limit on their
density at these redshifts, since we only selected sources
characterized by a well-defined stellar bump in the rest-frame
near-infrared.
This limit agrees with the predictions of the various models
dealing with infrared galaxy evolution 
\citep[e.g.,][]{Lagache04}.

Our selected sources exhibit a rather wide range of MIPS to IRAC flux
ratios and optical/NIR SED shapes. This suggests a possibly large
diversity in the properties of infrared galaxies at high redshift.
It is also interesting to note that the most distant sources
of our sample tend to
have extremely red colors from the $B$ to the 24\mic band,
with a global SED best fitted by ULIRG templates such as the
one from Arp\,220.
Finally, our data suggest the presence at high redshift
of the broad-band emission features often seen in 
the mid-infrared spectra of local galaxies and commonly attributed
to large molecules transiently heated by UV photons.
Even though we can not completely exclude 
 a  steeply rising continuum accounting for the
24\mic emission, the observed MIPS/IRAC flux
ratios are indeed difficult to reproduce without a 
significant contribution from these features (see Fig.\,2). A similar
conclusion is  reached by \citet{Lagache04} based
on the modeling of the MIPS 24/70/160\mic source number counts.

Since the late 1990's, there have been 
a wealth of 
discoveries at very high redshift (e.g., $z \gtapp 3$), while
little progress has been made in the 
 $1 \ltapp
z \ltapp 2$ redshift range \citep[but see][]{Steidel04}.
Yet, galaxy evolution at $1 \ltapp z \ltapp 2$ played a central role
in the cosmic growth of structures and the assembly of galaxies
\citep{Ellis01a}. MIPS
provides a unique opportunity to explore the star
formation and black hole accretion history at these redshifts.
Compared to previous models for instance, the new predictions by
\citet{Lagache04} based on the MIPS source number counts  indicate
a high density of $z$$\sim$1.5 LIRGs/ULIRGs, the first of which have
just been identified in this paper.
The MIPS deep surveys
 will reveal
the evolution of luminous and massive galaxies at
$1 \gtapp z \gtapp 3$, bridging the gap between the different
populations revealed by $ISO$ and SCUBA. Combining
 with sources at similar redshifts but selected with
different techniques (e.g., Balmer-break galaxies) will also
give us new clues on galaxy evolution.

\acknowledgments We ackowledge Rob Ivison and an anonymous
referee for their helpful comments on the manuscript.
We thank the funding from the MIPS and IRAC projects,
 which are both supported by NASA through the Jet Propulsion Laboratory, 
subcontracts
\#960785 and \#1256790. 
This work also  uses data obtained with support of the
National Science Foundation grants AST 95-29028 and AST 00-71198
awarded to S.M.Faber. Additional $K$-band data was kindly made available
to us from observations of the EGS and other DEEP2 fields 
being
undertaken at the Hale 5m Telescope by C.\,Conselice, K.\,Bundy and
R.\,Ellis.
We are indebted to the
Instituto de Astrof\'{\i}sica de Canarias for maintaining
its access to the WFC archival data. We
are also particularly grateful to Jim Cadien for helping us in
the data analysis process, and we acknowledge Guilaine
Lagache for the access to
her model predictions.

\clearpage

\begin{figure}
\plotone{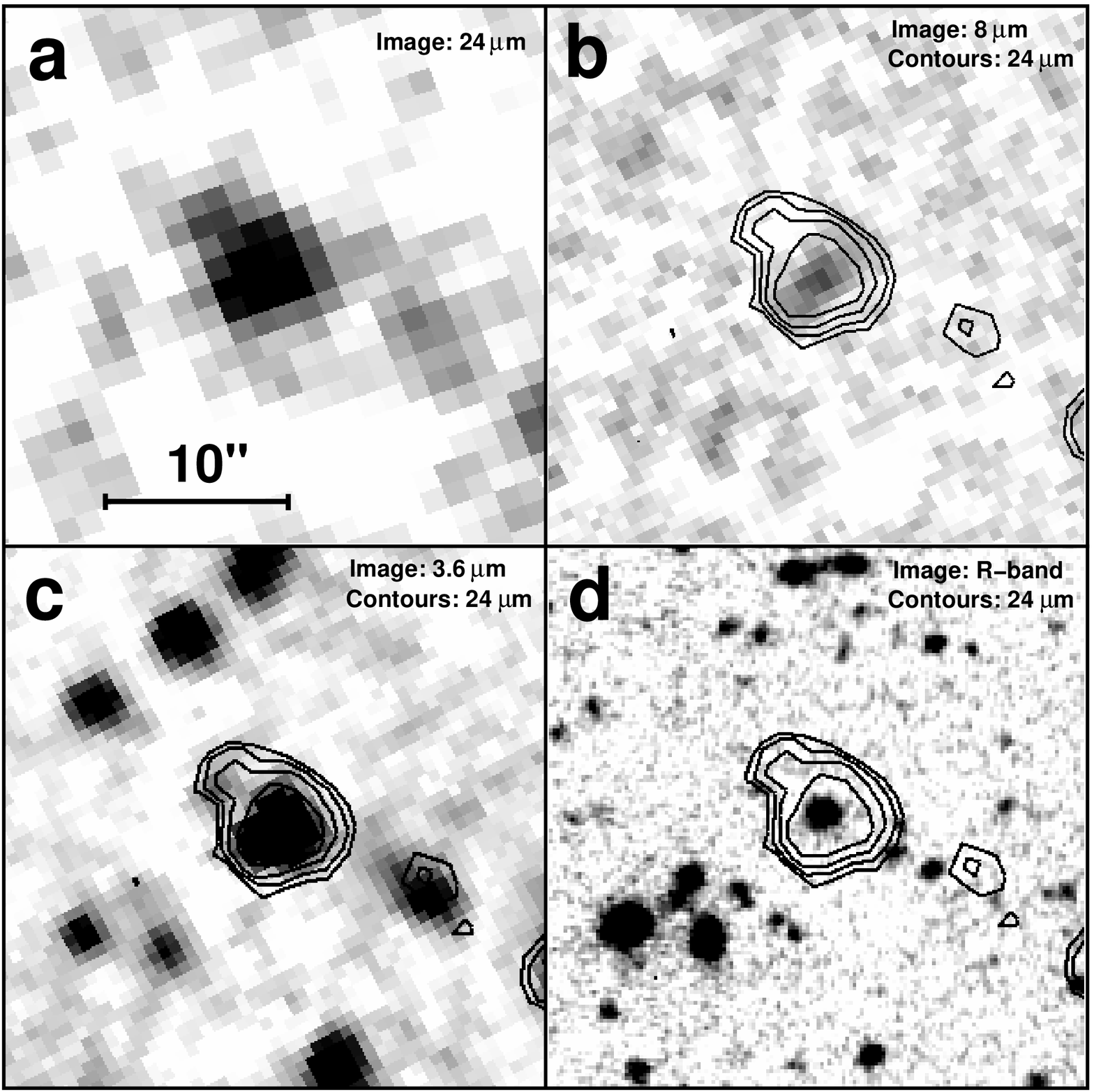}
\caption{
An illustration of our band-merging process, showing a 
source detected with MIPS at 24\mic (a)
and its  counterparts at 8\mic (b), 
3.6\mic (c) and 0.6\mic in the $R$--band (d).
Contours in these 3 last panels
correspond to the 24\mic emission. 
The centering accuracy at this wavelength ($\ltapp 2''$), combined with
the sharp PSF of the IRAC data between 3\mic and 9\mic allow  
unambiguous identifications. The four  images have
a similar field of view, with the scale indicated in the top-left
panel. We estimate for this source a photometric redshift $z$$\sim$1.8 
(see Sect.\,3). 
}
\label{fig:band_merging}
\end{figure}

\begin{figure}
\begin{center}
\includegraphics[width=11cm]{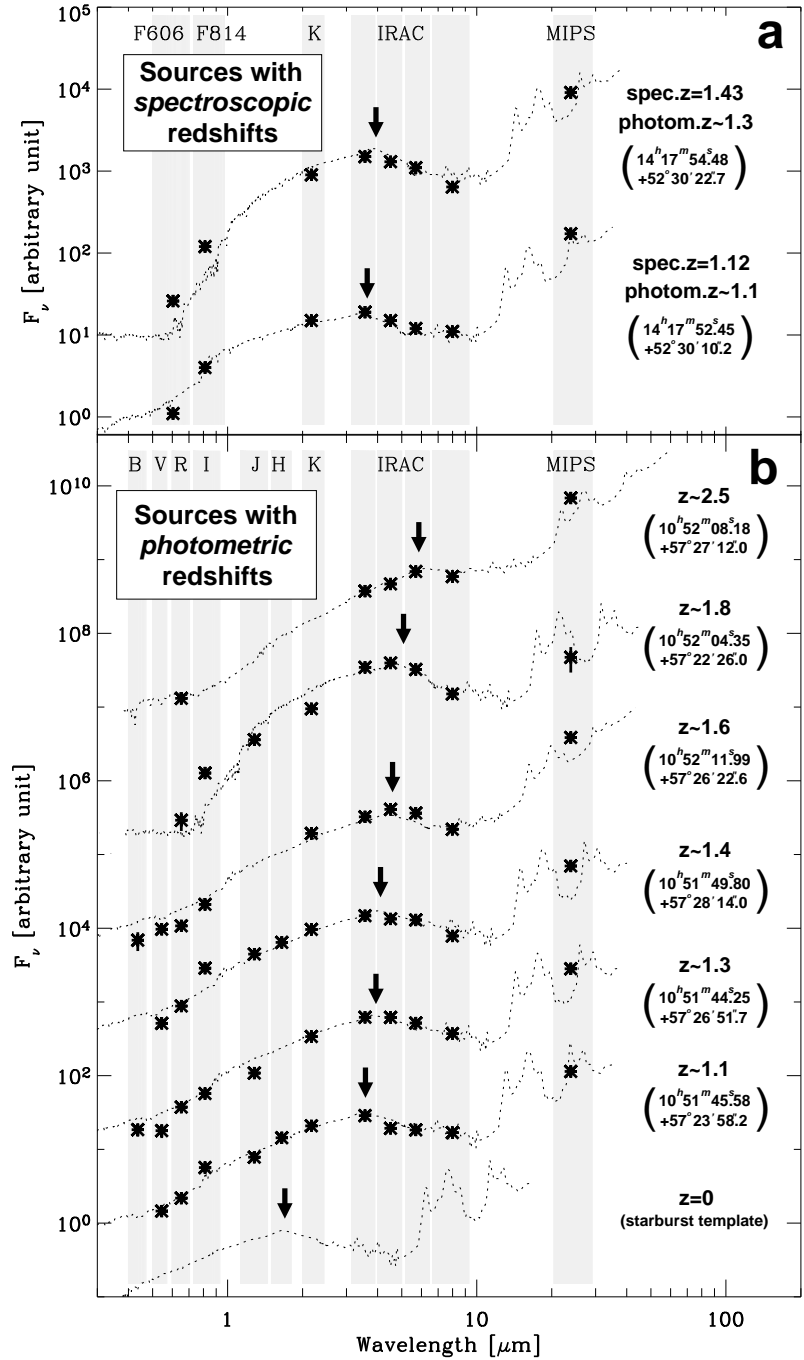}
\caption{ Optical/NIR/IRAC/MIPS broad-band photometry
('$\ast$' symbols) of 24\mic selected sources
with spectroscopic (a) and photometric (b) redshits $z \gtapp 1$.
In each case, the dashed
curve represents the best redshifted SED template fitting the data
(see Sect.\,\ref{sec:temp}
 for more details). The corresponding photometric redshift and the
source coordinates (J2000 epoch) are 
mentioned to the right of the panel. 
Flux uncertainties are represented with vertical solid lines.  A $z$=0
 starburst-like spectrum is added at the bottom for reference.
Note how the rest-frame 1.6\mic stellar bump gives a
clear constraint on the redshift determination ($\Downarrow$ symbol).  }
\vskip .2cm
\label{fig:SED}
\end{center}
\end{figure}

%
   
\clearpage

\begin{deluxetable}{lcccc}
\singlespace
\tabletypesize{\footnotesize}
\tablewidth{0pt}
\tablecaption{General properties of our selected sources in the Lockman Hole
\label{tab:conf}}
\tablehead{
\colhead{Redshift} &
\colhead{type$^a$} &
\colhead{N$_{\rm o}$$^b$} &
\colhead{L$_{\rm IR}$(10$^{11}$L$_{\odot}$)$^c$} &
\colhead{M$_{\rm K}$$^{c,d}$} 
}
\startdata
\vspace{.1cm}	     									      
1.0 -- 1.5  & Red Sp. & $\geq$3  & 4$_{\,3}^{\,6}$  & $-25.2_{-25.6}^{-24.8}$ \\
\vspace{.1cm}	     									      
            & Starb.  & $\geq$11 & 6$_{\,1}^{\,10}$ & $-24.8_{-25.7}^{-23.8}$ \\
\vspace{.2cm}	     									      
            & VLIRG   & $\geq$3  & 18$_{\,11}^{\,22}$ & $-25.7_{-26.3}^{-24.9}$ \\
\vspace{.1cm} 	     									      
1.5 -- 2.0  & Red Sp. & $\geq$1  & 4                        & $-26.4$           \\
\vspace{.2cm}	     									      
            & VLIRG   & $\geq$7  & 26$_{\,8}^{\,55}$ & $-25.4_{-27.0}^{-24.6}$ \\
2.0 -- 2.5  & VLIRG   & $\geq$3  & 58$_{\,20}^{\,90}$  & $-26.0_{-26.4}^{-25.4}$ \\
\enddata
\tablenotetext{a}{Source classification -- Red Sp.: IR-quiescent early-type spiral galaxy with a red optical/NIR color. 
Starb.: M82-like IR starburst associated with a late-type spiral. VLIRG: Very Luminous IR Galaxy characterized
by a high rest-frame IR/NIR ratio and a rather elusive 4000\AA \, break.}
\tablenotetext{b}{Minimum number of selected sources per category.}
\tablenotetext{c}{assuming H$_0$\,=\,70~km~s$^{-1}$\,Mpc$^{-1}$,
$\Omega_m$\,=\,0.3 and $\Omega_{\lambda}\,=\,0.7$. The mean value is given, along with the range of 
measurements.}
\tablenotetext{d}{Rest-frame $K$-band absolute magnitude.}
\end{deluxetable}

\end{document}